\documentclass{aa}
\usepackage[dvips]{graphicx}
\usepackage{natbib}
\bibpunct{(}{)}{;}{a}{}{,}
\begin{document}

\title{Reprocessing the {\it Hipparcos} data for evolved giant stars II. 
Absolute magnitudes for the R-type carbon stars\thanks{Based on 
observations from the Hipparcos
astrometric satellite operated by the European Space Agency (ESA 1997)}}
\titlerunning{Carbon R star absolute magnitudes} 

\author{G.~Knapp\inst{1}\and D.~Pourbaix\inst{1,2}\fnmsep
\thanks{Post-doctoral Researcher, F.N.R.S., Belgium}\and A.~Jorissen
\inst{2}\fnmsep\thanks{Research Associate, F.N.R.S., Belgium}} 
\institute{Department of Astrophysical Sciences, Princeton University, 
Princeton, NJ 08544-1001, U.S.A.\and Institut d'Astronomie et d'Astrophysique, 
Universit\'e Libre de Bruxelles, C.P.~226, Boulevard du Triomphe, B-1050 
Bruxelles, Belgium}
\date{Received date / Accepted date} 
\offprints{G.~Knapp \email{gk@astro.princeton.edu}}

\abstract{
The Hipparcos {\it Intermediate Astrometric Data} for carbon stars have been 
reprocessed using an algorithm which
provides an objective criterion for rejecting anomalous data points
and constrains the parallax to be positive.
New parallax solutions have been derived for 317 cool carbon stars, mostly
of types R and N.  In this paper we discuss the results for the R stars.
The most important result is that the early R stars (i.e., R0 
-- R3) have absolute magnitudes and $V-K$ 
colors locating them among red clump giants in the Hertzsprung-Russell
diagram. 
The average absolute magnitude $M_{\rm K}$ for early R-type stars
(with $V - K < 4$) has been derived from a Monte-Carlo simulation
implicitly incorporating all possible biases. It appears that the simulated
magnitude distribution for a population with a true Gaussian distribution of
mean $M_K = -2.0$ and intrinsic standard deviation 1.0~mag
provides a satisfactory match to the observed distribution.  
These values are consistent with the average absolute magnitude  
$M_{\rm K} = -1.6$ for clump red giants in the solar neighborhood 
\citep{Alves-2000:a}.
Further, early R-type stars are non-variable, and 
their infrared photometric properties show that they are not undergoing
mass loss, properties similar to those of the red clump giants.\\
Stars with subtypes R4 -- R9 tend to be cooler and have similar luminosity to
the N-type carbon stars, as confirmed by their position in the 
$(J-H, H-K)$ color-color diagram.\\
The sample of early R-type stars selected from the Hipparcos Catalogue appears
to be approximately complete to magnitude $K_0 \sim 7$, translating into a 
completeness distance of 600~pc if all R stars had $M_K= -2$ (400~pc if 
$M_K= -1$). With about 30 early R-type stars in that volume, 
they comprise about 0.04\% (0.14\%\ for $M_K= -1$) of the red clump stars in the
solar neighborhood. Identification with the red clump locates these
stars at the helium core burning stage of stellar evolution,
while the N stars are on the asymptotic giant branch, where
helium shell burning occurs.  
The present analysis suggests
that for a small fraction
of the helium core burning stars (far lower than the fraction of helium
shell-burning stars), carbon produced in the interior is mixed to 
the atmosphere in sufficient quantities to form a carbon star.
}

\maketitle

\section{Introduction}

The Harvard spectroscopic classification of stars for the Henry Draper
Catalogue recognized the first examples of chemically-peculiar stars, the 
carbon and S giants.   Carbon stars are immediately recognizable by the
presence of absorption bands of $\rm C_2$ at 4383 \AA{} and 
4737 \AA{} and the 4216 \AA{} band of CN \citep{Shane-1928:a},
and are chemically characterized by having C/O $>$ 1.
\citet{pickering-1896:a,Pickering-1908:a} 
showed that the carbon stars further subdivide
into two spectral classes, the `warm' R stars and the `cool' N stars,
and in the intervening years it has become increasingly apparent that these
stars do not represent a continuum of properties but rather belong to
two distinct classes. Further,
\citet{Keenan-1942:a} noted that some members of the R spectral
class were metal-poor and had high radial velocities and proper
motions.  These are now separately classified as CH stars.  Several 
low-concentration globular clusters contain CH stars 
\citep{Bond-1975:a,Mcclure-1984:b}; they are generally considered to be 
halo carbon stars.  The most recent list of
CH stars is by \citet{Bartkevicius-1996:a}.
Several additional types of carbon star have been 
recognized, including R~CrB stars and dwarf carbon stars.

The revised classification system of \citet{Keenan-1993:a}
is an attempt to provide consistent temperature, luminosity and abundance
sequences among the different types of carbon stars.
Three main classes are defined in this scheme: C-R$n$, C-N$n$, and 
C-H$n$, corresponding to the old R, N and CH types respectively,
with $n$ being a homogeneous temperature index for the three sequences,
corresponding to the G-K-M temperature sequence in ``oxygen'' stars.
Unfortunately, spectral types based on this scheme are available for only 
a small fraction of carbon stars.

Since C/O $< 1$ everywhere in the interstellar medium, in the Sun,
and in most stars, the excess carbon in carbon stars must be the result of 
stellar nucleosynthesis in the star itself or in a binary companion.
The many varieties of carbon stars are thus stars in the late stages of
evolution or with evolved binary companions, and are preferentially
formed in systems whose intrinsic 
metallicities (including the oxygen abundance)
are low.

The N stars are understood in a general sense.  They are very luminous, 
cool stars with shell hydrogen and helium burning 
and with 3$\alpha$-produced 
$\rm ^{12}C$ mixed to the surface by deep convection \citep{Iben-1983:a}, 
although the details of how that mixing occurs are far
from well understood.  Other stellar types with enhanced carbon abundance,
like CH and barium stars, are observed to be members of binary systems 
\citep[see the discussions by][]{Mcclure-1984:a,Mcclure-1997:a,Jorissen-1998:a},
and their peculiar abundances
can be attributed to mass transfer.  The R CrB stars are 
variable hydrogen-deficient stars, 
and have shed most of their envelope.  The dwarf carbon stars \citep{Dahn-1977:a,Green-1994:a} and the R stars remain a mystery.
Like the N stars, the R stars are red giants, but their average luminosity,
determined by statistical parallax methods \citep{Vandervort-1958:a,Scalo-1976:a,Dean-1976:a} is too low for shell helium burning.
Further, their oxygen abundances are similar to that of the Sun \citep{Dominy-1984:a}, they do not exhibit s-process overabundances, and 
no R star has been found so far in a binary system \citep{Mcclure-1997:b}, 
a statistically unlikely and surprising result.
It has therefore been suggested that R stars might actually be
coalesced binary systems.
The remaining hypothesis is that these stars are helium core-burning stars or 
post helium core burning stars, with carbon produced by helium burning
somehow introduced into their atmospheres, perhaps at the helium 
core flash \citep{Scalo-1976:a,Dominy-1984:a}.  There is little
direct evidence for this hypothesis, though, because of lack of information
about the intrinsic luminosity function of the R stars or of their spatial
density.  This paper sets out to derive the luminosities of the R stars
to attempt to discover their location on the Hertzsprung-Russell diagram
and establish their evolutionary status.

The Hipparcos project \citep{Hipparcos}
made it possible for the first time to make direct estimates of the distances
to giant carbon stars.  However, as noted by 
\citet{Knapp-2001:a}, the parallax errors for these stars are often
far larger than expected.  In this paper, we use new parallaxes derived
from a re-reduction of the Hipparcos {\it Intermediate Astrometric Data}
to show that essentially 
all the (early) R stars have the same absolute magnitude and color,
and are located in the red clump region of the HR diagram.  
The Hipparcos parallaxes
also allow us to derive the spatial density of the red clump stars and
the fraction of these which are R stars; we show that they
are very rare, far less 
than 1\% of the red clump stars (Sect.~4).  The carbon enhancements in R
stars may therefore originate in the helium-burning interiors, 
perhaps as the result of a mixing in a rapidly rotating
  star spun up by the accretion of its companion (see Sect.~\ref{Sect:conclusions}).
Sect.~2 discusses the astrometric, photometric,
spectroscopic and variability data on which the results presented in this
paper are based.  The absolute magnitudes and colors
are examined in Sect. 3, the space densities in Sect. 4,
and the implications and conclusions are discussed in Sect. 5. 
The details of the re-reduction of the Hipparcos data will be discussed
by \citet[][Paper {III}]{Pourbaix-2001:c} 
and the results for N-type
carbon stars by \citet[][Paper {IV}]{Pourbaix-2001:d}.

\section{The data}

\subsection{Overview}
\label{Sect:overview}

The basic data for R stars, including the new parallax data,
are collected in Tables~\ref{Tab:dataearly} \& \ref{Tab:datalate}.
Stars with enhanced carbon abundances come in many spectral types: the
N stars,
the R stars, carbon white dwarfs, carbon-rich Wolf-Rayet stars,
carbon cepheids, R CrB stars, barium stars, CH stars, and dwarf carbon
stars, among others.  As pointed out above, their chemical peculiarities are
due to nucleosynthesis within the star itself, or its companion, and the
task is to identify chemical peculiarities with particular evolutionary
paths. 
This is carried out in the first instance
by associating like objects, classifying them by spectral type, luminosity,
color, and variability, and is made difficult both by the wide variety
of stars with carbon enhancements, and by the often uncertain observational
information. 
A given star can be classified as several different
spectral types in several different papers, causing confusion in attempts 
to evaluate the common properties
of given classes of stars \citep[see the
discussions by][]{Mcclure-1997:a,Mcclure-1997:b,Wallerstein-1998:a}.  
Examples include  HIP~12028 (HD~161115) which is
classified as R by \cite{Vandervort-1958:a}, as CH by \cite{Keenan-1993:a} and as J by \cite{Ohnaka-1999:a}, while HIP~85750
(BD$+02^\circ3336$) is classified as R2 by \cite{Vandervort-1958:a},
as CH by \cite{Yamashita-1975:a}, and as N by \cite{Barnbaum-1996:a}.  
Thus it is far from straightforward to decide, from the available data,
whether the presence of a star in a given variability class with a spectral
type atypical of that class demonstrates a range of spectral types in that
variability class or is due to inadequate data. The infrared
color-color diagram $(J-H, H-K)$ (Fig.~\ref{Fig:JHK}) may nevertheless 
be of some help in the process of identifying misclassified stars, as
the different types of carbon stars occupy specific locations in that
diagram (see Sect.~\ref{Sect:photom}).

\begin{table*}[htb]
\caption[]{\label{Tab:dataearly}Early R stars: Astrometric, 
photometric and spectroscopic
  data. The rederived parallax $\varpi_{\circ}$ and its lower and upper
bounds (respectively  $\varpi_L$ and $\varpi_U$; see text) 
are given in milliarcseconds (mas). The R stars have been separated
into early and late types, according to the $V-K$ color index being
smaller or larger than 4 (see Fig.~\protect\ref{fig1}). 
}
\begin{tabular}{llllllllllllllll}
\hline
HIP & GCGCS & $\varpi_L$ & $\varpi_{\circ}$ & $\varpi_U$ & $V$ & $K$ & $J-H$ &
$H-K$ & Ref & Sp.& Ref    & Rem. \\
    &      &   (mas)    &  (mas)           & (mas)      &     &     &       & 
      & $JHK$&type& SpT\\   \hline
  2700 & 80   &  1.32 &  2.37 &  4.24 &  9.62 & 6.94 & 0.51 & 0.16 & D86 & R2 & B96 \\ 
  5809 & 196  &  0.24 &  1.09 &  5.05 & 10.07 & 7.22 & 0.77 & 0.03 & C79 & R0 & HD \\ 
  7816 & 256  &  --   &  --   & --    & 11.11 & 7.90 & 1.02 & 0.17 & D86 & R0 & S69 \\ 
 11508 & 357  &  0.25 &  1.39 &  3.78 &  9.57 & 5.92 &  --  &  --  & D86 & R0 & V58 \\ 
 12028 & 378  &  3.52 &  4.57 &  5.93 &  8.15 & 5.38 & 0.62 & 0.15 & F92 & R2 & V58 \\ 
 17933 & 576  &  0.54 &  1.11 &  2.29 &  8.29 & 4.65 & 0.73 & 0.19 & C81 & R3 & HD \\ 
 18564 & 588  &  0.37 &  1.49 &  4.51 & 10.26 & 6.66 & 0.41 & 0.23 & D86 & R3 & S28 \\ 
 18696 & 594  &  --   &  --   &  --   & 10.45 & 7.23 & 0.58 & 0.15 & 2M & R2 & S28 \\ 
 19050 & 610  &  --   &  --   &  --   & 10.71 & 8.10 & 0.48 & 0.16 & 2M & R0 & V58 \\ 
 19269 & 639  &  0.72 &  1.74 &  4.20 & 10.62 & 6.97 & 0.56 & 0.46 & C79 & R3 & HD \\ 
 26927 & 1035 &  0.17 &  1.08 &  6.57 & 10.02 & 7.62 & 0.5 & 0.09 & D86 & R0 & V58 \\ 
 28172 & 1110 &  0.27 &  1.33 &  4.40 & 10.35 & 7.62 & 0.53 & -0.11 & D86 & R2 & S28 \\ 
 31725 & 1380 &  0.46 &  1.10 &  2.66 &  9.41 & 6.46 & 0.54 & 0.21 & C79 & R0 & HD \\ 
 33042 & 1460 &  --   &  --   &  --   & 10.56 &  --  &  --  &  --  &  --  & R & S72 \\ 
 35681 & 1622 &  0.37 &  1.05 &  2.99 &  8.52 & 6.16 & 0.4 & 0.18 & 2M & R0 & V58 & *\\ 
 36086 & 1703 &  0.13 &  1.08 &  4.48 &  9.22 & 5.81 & 0.61 & 0.34 & D86 & R3 & S28 \\ 
 39118 & 1981 &  0.06 &  1.32 &  7.23 & 10.41 & 7.23 & 0.79 & 0.25 & 2M & R2 & V58 \\ 
 40374 & 2126 &  --   &  --   &  --   & 11.70 & 8.38 & 0.6 & 0.22 & 2M & R & GCGCS \\ 
 44172 & 2396 &  0.81 &  2.08 &  5.34 &  9.44 & 6.84 &  --  &  --  & D86 & R2 & B96 \\ 
 48329 & 2626 &  0.35 &  1.22 &  4.23 &  9.90 & 6.25 & 0.7 & 0.22 & D86 & R3 & S28 \\ 
 50994 & 2759 &  0.35 &  1.05 &  3.14 &  9.56 &  --  &  --  &  --  &  --  & R0 & HD \\ 
 53354 & 2892 &  --   &  --   &  --   & 10.54 & 7.62 & 0.61 & 0.17 & 2M & R2 & V58 \\ 
 53522 & 2900 &  0.63 &  1.81 &  5.15 & 10.14 &  --  &  --  &  --  &  --  & R0 & V58 \\ 
 53810 & 2925 &  --   &  --   &  --   &  8.35 &  --  &  --  &  --  &  --  & K5R & S44 \\ 
 53832 & 2919 &  0.96 &  2.08 &  4.52 & 10.11 &  --  &  --  &  --  &  --  & R0 & V58 \\ 
 56405 & 3058 &  0.12 &  1.19 &  6.52 & 10.27 &  --  &  --  &  --  &  --  & R2 & S28 \\ 
 58786 & 3156 &  0.26 &  1.06 &  4.20 & 10.27 & 7.51 & 0.52 & 0.15 & 2M & R2 & V58 \\ 
 62944 &      &  7.41 &  8.22 &  9.12 &  6.91 & 4.16 & 0.55 & 0.1 & E78 & R3 & B96 \\ 
 63955 & 3335 &  1.44 &  2.26 &  3.54 &  8.50 & 5.97 & 0.46 & 0.18 & 2M & R3 & B96 \\ 
 65320 & 3379 &  1.84 &  2.94 &  4.71 &  9.66 &  --  &  --  &  --  &  --  & R0 & S28 \\ 
 66317 & 3409 &  --   & --    & --    & 12.39 &  --  &  --  &  --  &  --  & R & S72 \\ 
 68543 & 3469 &  0.21 &  1.02 &  5.07 &  9.48 & 6.67 & 0.52 & 0.19 & D86 & R2 & S28 \\ 
 69089 &      &  0.53 &  1.17 &  2.59 &  8.68 & 6.40 & 0.46 & 0.12 & 2M & R2 & B96 \\ 
 71464 & 3513 &  --   & --    & --    & 12.17 &  --  &  --  &  --  &  --  & R & S72 \\ 
 73955 & 3591 &  0.42 &  1.17 &  3.28 &  9.99 &  --  &  --  &  --  &  --  & R0 & S28 \\ 
 74179 & 3562 &  0.28 &  1.16 &  4.83 &  9.72 & 6.50 & 0.5 & 0.5 & F97 & R3 & HD & *\\ 
 74826 & 3586 &  0.52 &  1.55 &  4.64 &  9.78 &  --  &  --  &  --  &  --  & R0 & S28 \\ 
 75691 & 3614 &  --   & --    & --    & 12.88 & 11.12 & 0.4 & 0.06 & 2M & R5 & V58 & *\\ 
 75745 & 3605 &  --   & --    & --    & 11.79 &  --  &  --  &  --  &  --  & R & S72 \\ 
 80365 & 3687 &  --   & --    & --    & 11.18 & 8.16 & 0.23 & 0.5 & F73 & R & M51 & *\\ 
 80769 & 3704 &  0.37 &  1.37 &  5.11 & 10.40 & 7.50 &  --  &  --  & M65 & R2 & V58 \\ 
 82184 & 3735 &  0.76 &  1.74 &  3.99 &  9.10 & 6.46 & 0.57 & 0.12 & 2M & R2 & B96 \\ 
 84266 & 3795 &  2.30 &  2.89 &  3.62 &  7.60 & 5.11 & 0.3 & 0.23 & N81 & R2 & B96 \\ 
 85117 & 3816 &  1.88 &  3.33 &  5.91 &  9.57 &  --  &  --  &  --  &  --  & R3 & C21 \\ 
 86927 & 3879 &  3.77 &  4.78 &  6.05 &  8.71 & 6.14 & 0.53 & 0.14 & D86 & R0 & S28 \\ 
 87603 & 3912 &  0.30 &  1.05 &  3.66 & 10.72 & 7.92 & 0.55 & 0.18 & 2M & R4 & S44 \\ 
 88584 & 3938 &  2.07 &  3.49 &  5.88 &  9.82 &  --  &  --  &  --  &  --  & R3 & S44 & *\\ 
 89239 & 3973 &  4.56 &  6.37 &  8.91 & 10.82 & 7.41 & 0.61 & 0.24 & 2M & R5 & V58 \\ 
\hline
\end{tabular}
\end{table*}

\addtocounter{table}{-1}
\begin{table*}[htb]
\caption[]{(cont.)}
\begin{tabular}{llllllllllllllll}
\hline
HIP & GCGCS & $\varpi_L$ & $\varpi_{\circ}$ & $\varpi_U$ & $V$ & $K$ & $J-H$ & $H-K$ & Ref & Sp.& Ref    & Rem. \\
    &      &   (mas)    &  (mas)           & (mas)      &     &     &       &      & $JHK$&type& SpT\\   \hline
 89739 & 3982 &  2.53 &  4.61 &  8.43 & 10.67 & 8.00 & 0.5 & 0.5 & F97 & R0 & C21 & *\\ 
 90199 & 4002 &  --   &  --   & --    & 11.91 &  --  &  --  &  --  &  --  & R0 & S44 \\ 
 90694 & 4021 &  0.15 &  1.19 &  5.56 &  9.88 & 6.30 & 0.67 & 0.23 & C79 & R5 & S44 \\ 
 92207 & 4098 &  0.26 &  1.62 &  6.91 & 10.00 & 6.5 & 1 & 1 & F97 & R0 & HD & *\\ 
 93987 & 4181 &  0.33 &  2.09 & 13.40 & 11.08 & 5.89 & 0.54 & 0.52 & 2M & R2 & S44 & *\\ 
 95392 & 4485 &  0.39 &  1.06 &  2.86 &  9.66 &  --  &  --  &  --  &  --  & R2 & S28 \\ 
 98057 & 4560 &  0.24 &  1.12 &  5.23 &  9.62 &  --  &  --  &  --  &  --  & R0 & S28 \\ 
 99725 & 4784 &  0.48 &  1.10 &  2.49 &  9.75 & 6.05 & 0.7 & 0.15 & U97 & R0 & V58 \\ 
102726 & 4972 &  --   & --    & --    & 10.64 &  --  &  --  &  --  &  --  & R1 & S44 \\ 
105212 & 5227 &  0.98 &  2.31 &  5.44 &  9.71 &  --  &  --  &  --  &  --  & R2 & V58 \\ 
105241 & 5230 &  1.56 &  3.11 &  6.19 &  9.81 & 6.97 & 0.5 & 0.15 & D86 & R2 & V58 \\ 
107349 & 5408 &  0.20 &  1.04 &  5.42 & 10.16 & 5.51 & 0.91 & 0.36 & C79 & R0 & HD \\ 
114452 & 5822 &  0.72 &  1.70 &  4.00 &  9.99 & 7.03 & 0.47 & 0.2 & D86 & R2 & S28 \\ 
114509 & 5823 &  1.92 &  3.24 &  5.47 &  9.30 & 6.69 & 0.96 & 0.11 & D86 & R0 & S44 \\ 
117467 & 5937 &  0.98 &  1.97 &  3.95 &  8.52 & 5.42 & 0.61 & 0.21 & C79 & R3 & B96 \\ 
\hline
\end{tabular}
\\
{\bf Notes}\\
HIP 35681: RU Cam, type CWa, period $22$ d;
HIP 74179: S Aps, type R CrB;
HIP 75691: NSV 7110;
HIP 80365: RT Nor, type R CrB;
HIP 88584: W CrA, type SRb, period $125$ d;
HIP 89739: RS Tel, type R CrB;
HIP 92207: V CrA, type R CrB;
HIP 93987: SV Sge, type R CrB\\
{\bf References to spectral types and magnitudes:}\\
HD: Henry Draper Catalogue; 
GCGCS: \citet{Stephenson-1989:a}; 
2M: {\it Two Micron All-Sky Survey} \citep{Skrutskie-1997:a};
B96: \citet{Barnbaum-1996:a}; 
C21: \citet{Cannon-1921:a}; 
C79: \citet{Catchpole-1979:a};
C81: \citet{Cohen-1981:a};
D86: \citet{Dominy-1986:a};
E78: \citet{Elias-1978:a};
F73: \citet{Feast-1973:a};
F92: \citet{Feast-1992:a};
F97: \citet{Feast-1997:a}; 
M51: \citet{Mayall-1951:a}; 
M65: \citet{Mendoza-1965:a}; 
N81: \citet{Noguchi-1981:a};
S28: \citet{Shane-1928:a}; S44: \citet{Sanford-1944:a}; 
S69: \citet{Slettebak-1969:a}; S72: \citet{Stock-1972:a};
U97: \citet{Ulla-1997:a};
V58: \citet{Vandervort-1958:a}
\end{table*}

\begin{table*}[htb]
\tabcolsep 2pt 
\caption[]{\label{Tab:datalate}Late R stars: Astrometric, photometric
  and spectroscopic data.  Columns have the same meaning as in
  Table~\protect\ref{Tab:dataearly}. 
Additional columns provide the variability name, 
type and period (from GCVS), and the $K - [12] (= K + 2.5 \log
F12/28.3)$ color index (where $F12$ is the flux density in the IRAS 12
$\mu$m band expressed in Jy). The $K - [12]$ index is a mass-loss indicator,
non-mass-losing stars having $K - [12] \sim 0.7$ 
\citep[e.g.][]{Jorissen-1998:b}.}
\begin{tabular}{lllllllllllllllll}\hline
HIP & GCGCS & $\varpi_L$ & $\varpi_{\circ}$ & $\varpi_U$ & $V$ & $K$ &
$J-H$ & $H-K$ & Ref       & Sp. & Ref         & GCVS & Var & 
$P$ & $K - [12]$ & Rem\\
    &      & (mas)      & (mas)            & (mas)      &     &     &         
      &       & $JHK$    &  type        & SpT &     &             &      
  (d)      &\\
 \hline
 35810 & 1686 &  0.38 &  1.26 &  4.18 &  9.01 &  --  &  --  &  --  &  --  & R8 & S28 & V758 Mon & -- \\ 
 36623 & 1737 &  0.38 &  1.36 &  3.58 &  8.02 & 2.95 & 0.72 & 0.38 & N81 & R9 & S28 & NQ Gem & SR & 70 & 0.6 & * \\ 
 42975 & 2326 &  0.44 &  1.04 &  2.47 &  8.98 & 2.71 & 1.38 & 0.84 & W00a & R & S72 & R Pyx & Mira & 365 & 2.3 \\ 
 44812 & 2428 &  0.21 &  1.22 &  4.72 & 10.61 & 6.50 & 0.78 & 0.35 & 2M & R6 & S44 \\ 
 50412 & 2715 &  0.61 &  1.65 &  4.47 & 10.83 & 6.46 & 0.76 & 0.38 & 2M & R6 & S44 \\ 
 52656 & 2852 &  2.91 &  3.67 &  4.64 &  8.71 & 2.54 & 1.16 & 0.54 & K94 & R & K93 & TZ Car & SRb & 69 & 1.5\\ 
 54806 & 2975 &  --   &  --   &  --   & 10.14 &  --  &  --  &  --  &  --  & R5 & HD \\ 
 60534 & 3227 &  1.19 &  1.84 &  2.85 &  7.52 & 2.53 & 0.95 & 0.43 & C79 & R4 & B96 & S Cen & SR & 65 & 1.3 \\ 
 62401 & 3286 &  0.24 &  1.67 & 11.75 & 11.98 & 1.81 & 1.85 & 1.3 & W00a & R3 & S44 & RU Vir & Mira & 433 & 4.1 \\ 
 66070 & 3405 &  0.37 &  1.08 &  3.20 &  8.51 & 3.52 & 0.94 & 0.37 & C79 & R2 & B96 & V971 Cen & -- & -- & 1.1 \\ 
 85750 & 3842 &  0.39 &  1.43 &  4.00 &  9.42 & 5.15 &  --  &  --  & D86 & R2 & V58 \\ 
 91929 & 4086 &  0.37 &  1.58 &  5.62 &  9.75 &  --  &  --  &  --  &  --  & R4 & B96 & RV Sct & Lb & -- & --  \\ 
 94049 & 4179 &  --   & --    & --    & 10.39 &  --  &  --  &  --  &  --  & R4 & S28 \\ 
 95422 & 4263 &  0.37 &  1.45 &  5.69 & 11.01 & 6.74 & 0.75 & 0.26 & 2M & R5 & V58 \\ 
 98223 & 4567 &  0.83 &  2.25 &  6.10 &  9.36 & 5.15 &  --  &  --  & M65 & R8 & S28 \\ 
104522 & 5147 &  0.34 &  1.22 &  4.34 &  9.76 & 5.26 & 0.91 & 0.29 & C79 & R5 & HD & & & & 1.1\\ 
108205 & 5494 &  0.29 &  1.03 &  3.60 &  9.23 & 1.71 &  --  &  --  & IRC & R2 & V58 & LW Cyg & Lb & -- & 1.5 & *\\ 
109158 & 5577 &  0.46 &  1.55 &  5.25 & 10.12 & 2.57 & 1.21 & 0.75 & N81 & R & L44 & CT Lac & SRa & 555 & 2.4 \\ 
113150 & 5761 &  --   & --    & --    & 10.75 &  --  &  --  &  --  &  --  & R5 & V58 \\ 
113715 & 5791 &  0.77 &  1.69 &  3.72 &  9.66 & 2.99 & 1.02 & 0.82 & N81 & R8 & S44 & VY And & SRb & 149 & 1.7 \\ 
\hline
\end{tabular}
\\
{\bf Notes}\\
HIP 36623: also classified as symbiotic variable;
HIP 108205: Li star \citep{Boffin-1993:b}\\
{\bf References to spectral types and magnitudes:}\\
HD: Henry Draper Catalogue; 
IRC: \citet{IRC}; 
2M: {\it Two Micron All-Sky Survey} \citep{Skrutskie-1997:a};
B96: \citet{Barnbaum-1996:a}; 
C79: \citet{Catchpole-1979:a}; 
D86: \citet{Dominy-1986:a}; 
K93: \citet{Keenan-1993:a};
K94: \citet{Kerschbaum-1994:a}; 
L44: \citet{Lee-1944:a};
M65: \citet{Mendoza-1965:a}; 
N81: \citet{Noguchi-1981:a};
S28: \citet{Shane-1928:a}; 
S44: \citet{Sanford-1944:a}; 
S72: \citet{Stock-1972:a};
V58: \citet{Vandervort-1958:a};
W00a: \citet{Whitelock-2000:a}
\end{table*}

Of the many varieties of carbon star, the cool luminous carbon stars 
are of interest because they are expelling carbon-enhanced
mass into the interstellar me\-dium.
\citet{Knapp-2001:a} identified about 320 carbon stars in the Hipparcos 
catalogue.
An attempt was made to be complete (as far as the members of the Hipparcos
catalogue are concerned) for the luminous cool carbon stars, i.e., the R and
N stars (because these are the stars which are, or may be, losing mass).
Although the Hipparcos catalogue itself
is based only partly on a complete input catalogue (the Input Catalogue
contained both complete samples of stars -- designed as `Hipparcos
survey stars'\footnote{Only 3 such stars (HIP~12028, HIP~62944 and
HIP~84266 discussed herein) are present in the sample of R stars} --  and 
additional samples of stars of particular interest), it turns out to be 
reasonably complete for carbon stars to Hipparcos
magnitude $Hp ~ \sim$ 9.5 \citep{Knapp-2001:a} or to $K \sim 7$
(Fig.~\ref{Fig:space}). 

\subsection{Photometry}
\label{Sect:photom}

We use the $V$ magnitude given in the Hipparcos 
catalogue (field H5).  We acquired
$J, H$ and $K$ photometry from the literature, as listed in
Tables~\ref{Tab:dataearly} \& \ref{Tab:datalate} along with the bibliographic reference.  
This was found with the aid of
the electronic version of the literature compilation catalogue of \citet{CaInOb}.
In addition, we extracted $K_s$
photometry from the {\it Two Micron All Sky Survey} (2MASS) on-line
data release \citep{Skrutskie-1997:a}.  
In many cases, the stars are bright enough that they
are saturated in the 2MASS data and we use older photometry.  There
are no systematic offsets evident among the various photometric measures
for a given star, and the measurements (including the narrower 2MASS
$K_s$ band magnitudes) agree to better than about 0.2~mag in almost
all cases.  We were able to find $K$ magnitudes for all but 24 stars.
Although these observations were made at different epochs, R stars
are not in general variable (Sect.~\ref{Sect:varab}) 
and so there is little or no uncertainty
introduced into the $V$ or $K$ magnitudes by variability.

\begin{figure}
\resizebox{\hsize}{!}{\includegraphics{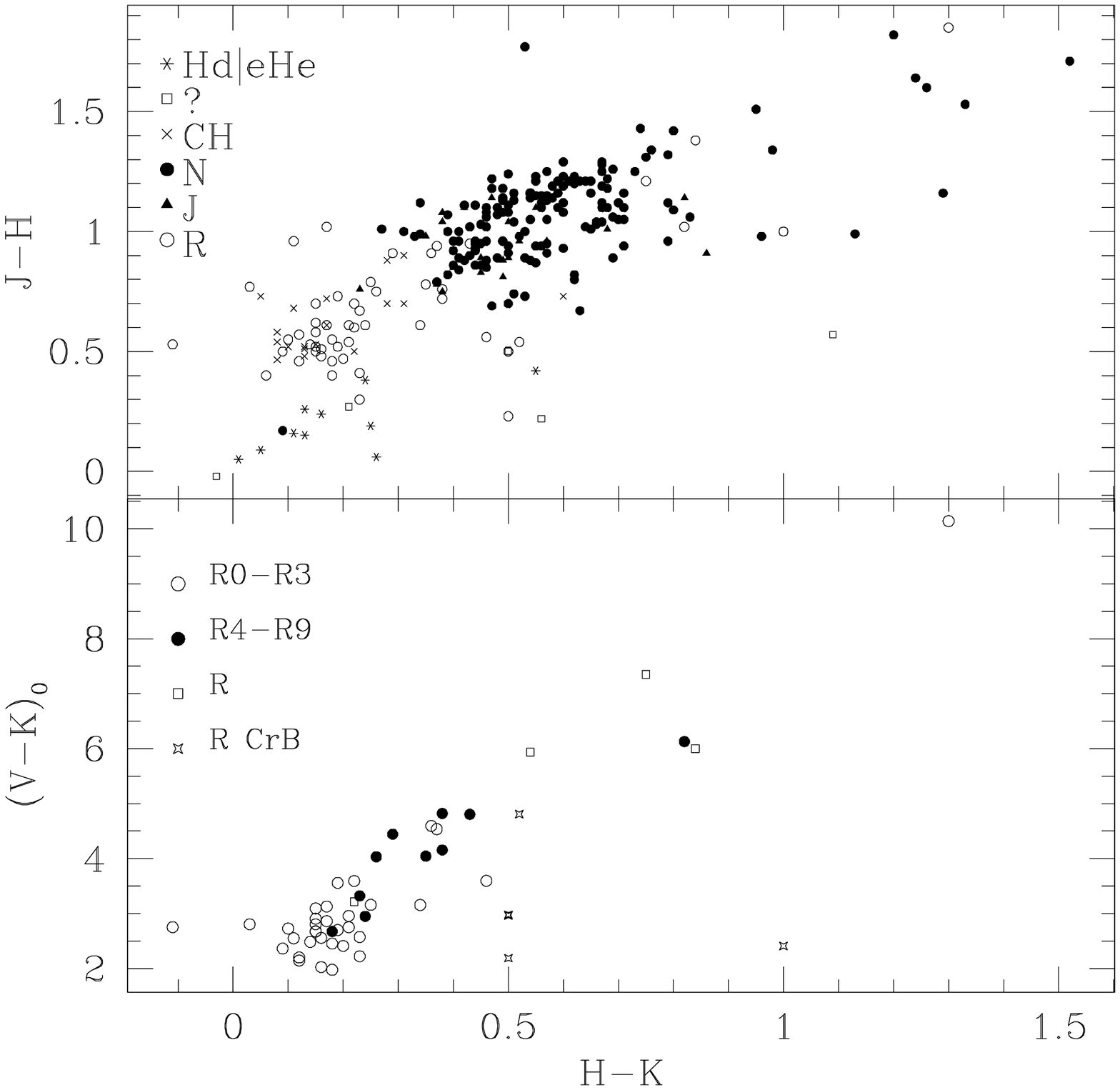}}
\caption[]{
Top panel: The $(J-H, H-K)$ color-color diagram for the carbon stars from the
Hipparcos catalogue with $J, H$ and $K$ collected from the literature 
(as listed in Tables~\ref{Tab:dataearly} \& \ref{Tab:datalate} for R stars). Symbols are as listed 
in the figure, with `Hd' and `eHe' standing for `hydrogen-deficient' and 
`extreme-helium' carbon stars. Hydrogen-deficient carbon stars are
similar to R CrB carbon stars, except that they do not undergo
obscuration episodes \citep{Feast-1997:a}. Extreme-helium carbon
stars are similar to, but hotter than, R CrB carbon stars \citep{Cottrell-1998:a}.\\
Bottom panel: The $((V-K)_0, H-K)$ diagram for the sample of R stars considered 
in the present study 
\label{Fig:JHK}
}
\end{figure}

Fig.~\ref{Fig:JHK} presents the $(J-H, H-K)$ color-color diagram. It
is immediately apparent that R stars occupy a very specific region of
that diagram, which is also populated by CH stars. A few R stars
appear to occupy the same region as N-type carbon stars, which is not
surprising as most -- if not all -- late R-type carbon stars may
actually be N stars (see the discussion in Sects.~\ref{Sect:varab}
and \ref{Sect:sptype}).

\subsection{Variability}
\label{Sect:varab}

N stars are in general long-period variables: irregular (Lb), 
semi-regular (SR or SRb) or Mira (Mira or SRa) --
see \citet{Kerschbaum-1994:a} and \citet{Wallerstein-1998:a}
for a discussion of the SRa and SRb stars.  R stars generally do not 
appear in variable star catalogues.  Of the 83 R stars in our sample,
18 appear in the {\it General Catalogue of Variable Stars} \citep{Kholopov-1998:a}, but most of these are late R stars, as listed in Table~\ref{Tab:datalate}.
Most of the R stars of variability
types Mira and SR are also losing mass, as indicated by their excess 12~$\mu$m
emission (Table~\ref{Tab:datalate}), 
and are therefore probably closely related to N stars. 

\subsection{Spectral Types}
\label{Sect:sptype}

There are two major studies of R stars. \citet{Vandervort-1958:a} observed a sample of 98 R stars selected from an
objective prism survey. \citet{Mcclure-1997:b} discusses a sample of
22 R stars. 
We have also used spectral types from \citet{Keenan-1993:a}, \citet{Barnbaum-1996:a}, and other
references. In toto, we have a sample of 83 R stars.  
The data 
are listed in Tables~\ref{Tab:dataearly} \& \ref{Tab:datalate}.  Where available, we have listed the
spectral type from R0 to R8. 

Many of the remaining 317 carbon stars are spectroscopically classified 
only as carbon stars, and some of these may be R stars also.
However, as the $(J-H, H-K)$ diagram shows (Fig.~\ref{Fig:JHK}), not
many non-R stars, apart from the CH stars, 
occupy the location of R stars in that diagram. 
CH stars must be treated separately from the R stars, as all CH stars are
believed to be binary systems \citep{Mcclure-1997:a}, and must
therefore be processed with a binary astrometric model, as presented in 
Paper~I \citep{Pourbaix-2000:b}. 

Further, since there is often disagreement in the various references as to the 
spectral type of a given star (examples were given in
Sect.~\ref{Sect:overview}), the `spectral purity' of the sample of
R stars collected in Tables~\ref{Tab:dataearly} \& \ref{Tab:datalate} is best assessed 
by looking at the $(J-H, H-K)$ diagram. 

The main source of confusion appears to be the fact that several late-type R stars
from Table~\ref{Tab:datalate} are located in the region occupied by N
stars. This is not totally surprising, as late R stars
bear similarities with, or might even be identical to, N
stars. In the following analysis, it will therefore be necessary to
treat separately R stars with $H-K < 0.3$ [or $(V-K)_0 < 4$; see
Fig.~\ref{Fig:JHK}] from those with $H-K \ge 0.3$.  
Finally, the few R stars also classified as R~CrB variables
should be treated separately as well.

To stress the fact that the early- and late-type R stars in the present sample
must be considered separately, they have been split into two separate
lists (Tables~\ref{Tab:dataearly} \& \ref{Tab:datalate}
respectively). The assignment of a star as an early- or late-type R star 
is mainly based on the $(V-K)_0$ index, with the threshold set
at 4. The $(V-K)_0$ index has preference over the R spectral subtype in case
of conflicting classifications. When the $(V-K)_0$ index is not available, a
criterion based on the spectral type is used (with the threshold set at R3.5). 
It will be shown below that, with this choice for the thresholds, the two
subsamples of R stars present homogeneous properties with respect to the R
spectral subtype, to mass loss  and to variability type, most late-type R stars
being long-period variable stars, with 
circumstellar dust emission indicative of mass loss. The $K - [12]$ color index
(with $K - [12] = K + 2.5 \log
F12/28.3$, $F12$ being the flux density in the IRAS 12
$\mu$m band, expressed in Jy) 
is considered as a good indicator of (dust) mass loss, according
to e.g., Fig.~21 of \citet{Jorissen-1998:b}. Non-mass-losing
stars have $K - [12] \sim$ 0.7.

Late-type R stars are thus likely to
belong to the mass-losing AGB, and
are therefore not considered further in the following discussion, which focuses on
R stars with $(V-K)_0 < 4$.

\subsection{Parallaxes}
\label{Sect:paral}

The Hipparcos catalogue \citep{Hipparcos} 
gives parallax values at the 
2$\sigma$ level or better [i.e. $\varpi/\epsilon(\varpi) >$ 2, where
$\epsilon(\varpi)$ is the quoted parallax uncertainty] for 17 R
stars, of which 15 are of types R0 -- R2.
The resulting absolute visual magnitudes 
for the early R stars cover a range of $M_{\rm V} = -2.8$ to +4.1 \citep{Knapp-2001:a}.  However, the parallax errors for many of the stars
are larger, sometimes much larger, than 
the errors expected for the magnitudes and ecliptic
coordinates of the stars. Of the nearly 320 carbon stars in the sample
of \citet{Knapp-2001:a}, only 25\% (and only 18\% of the R
stars) have ``2$\sigma$'' parallaxes.

The Hipparcos {\it Intermediate Astrometric Data} 
\citep[IAD;][]{vanLeeuwen-1998:a}
for the entire sample of carbon stars were therefore
reprocessed as described and Papers I \& III. 
Therefore, only a brief outline of the method is given here.

The position, distance modulus and proper motion are the five parameters in the
model fit to be adjusted by $\chi^2$ minimization of the IAD abscissa
residuals.  Included in this fitting procedure is the rejection of outliers, 
defined as having a fit residual greater than three times the average of the 
residuals.  In practice, two or three observations at most were rejected.
Fitting the distance modulus ensures that the derived parallax $\varpi$ is 
always positive, as described in Paper~I, but at the expense 
of asymmetric error bars and of a bias on $\varpi$. This bias is of
any importance only when  $\epsilon(\varpi)/\varpi \ga 1$, as shown in
Paper~III. The distance modulus is better behaved, and must be used 
for computing average
absolute magnitudes, as further discussed in Sect.~\ref{Sect:average}.

Another major difference with respect to the reduction scheme applied
by the Hipparcos consortia resides in the bootstrap algorithm 
used to reject those 
data points which yield inconsistent FAST and NDAC solutions, as
described in detail in Paper~III. 

\begin{figure}
\resizebox{\hsize}{!}{\includegraphics{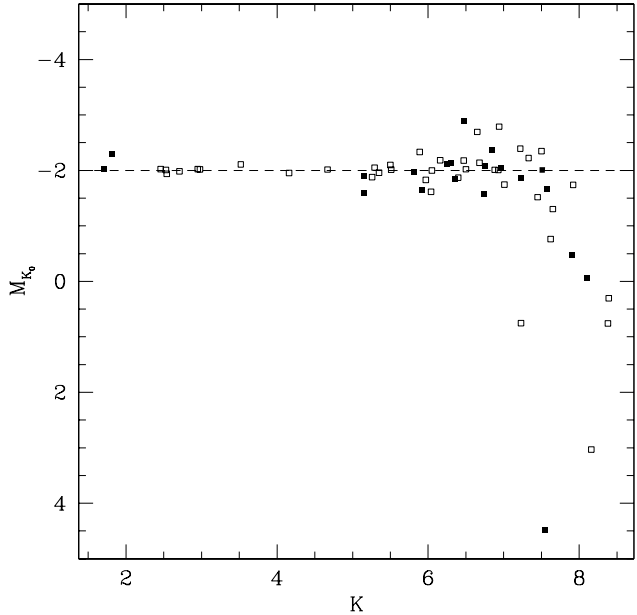}}
\caption[]{\label{Fig:MK_bias}
The absolute magnitudes $M_{\rm K}$ retrieved by processing the IAD 
as indicated in Sect.~\ref{Sect:paral} for the sample stars
assuming that they all have a true absolute magnitude $M_K^0
=  -2.0$ (see also Sect.~\protect\ref{Sect:subaverage}). 
It is  clearly seen that a strong bias on $M_K$ appears for stars with $K \ga
8$ that have noisy IAD. Stars with negative parallaxes in the
Hipparcos catalogue are identified by filled squares
}
\end{figure}

Monte-Carlo simulations have shown that this reprocessing scheme of the IAD
fails when the noise on the IAD is too large with respect to the true parallax,
in that it yields absolute magnitudes that are too faint by as much as 5
magnitudes (i.e., parallaxes come out too large). This is not really unexpected,
as the bootstrap method will remove `discrepant' observations until a solution
passing through the remaining IAD is finally found. The parallax of that
solution will clearly be of the order of the noise on the abscissa residuals,
though the spurious nature of that solution may be noticed by the large
uncertainty on the parallax. 
For the sample of R stars considered in the present
paper, such a bias on the absolute magnitude becomes clearly apparent
for stars with $K > 8$, although the parallax of 
somewhat brighter stars is already
biased as well, as 
seen in Fig.~\ref{Fig:MK_bias} (see also
Sect.~\ref{Sect:subaverage}). 
In an attempt to keep as many unbiased stars as possible, the
following criterion has been adopted to reject biased parallaxes: $K > 8$
(or $V > 10.5$ if no $K$ magnitude is available) or the star has a
simulated absolute magnitude that lies further than 4$\sigma$ away
from the true absolute magnitude of $-2$ on Fig.~\ref{Fig:MK_bias},     
$\sigma$ being the standard deviation of the simulated absolute
magnitudes for stars with $K < 7$ (i.e., which are not biased).
Therefore, Tables~\ref{Tab:dataearly} \& \ref{Tab:datalate} list only 
those parallaxes that may be considered as unbiased according to the
above criteria.

The results are listed in Tables~\ref{Tab:dataearly} \& \ref{Tab:datalate}, 
which provides the HIP number 
from the Hipparcos catalogue in the first column. The star number in the 
{\it General Catalog of Cool Galactic Carbon Stars} 
\citep[GCGCS;][]{Stephenson-1989:a} is listed in column 2.
Columns 3 to 5 provide the parallax $\varpi_0$ derived from the
$\chi^2$ minimization on the distance modulus, the corresponding
lower and upper bounds ($\varpi_L$ and $\varpi_U$) computed from Eq.~(9) of
Paper~I. The following columns list the 
apparent $V$ magnitude (field H5 of the Hipparcos
catalogue), the $K$ magnitude, the $J-H$ and $H-K$ color indices, 
their reference, and finally the spectral type and its reference.

\section{The absolute magnitudes and colors of R stars}
\label{Sect:average}

The absolute magnitudes $M_{\rm V}$ and $M_{\rm K}$
are calculated as follows:
\begin{equation}
d\;  = \;  1/\varpi('')
\end{equation}
\begin{equation}
M_{\rm X}\;  = \; m_{\rm X} + 5 -5 \log d - A_{\rm X} \qquad \mbox{\rm 
  (where}\;
$X = V$\; {\rm or}\; $K$)
\end{equation}
\begin{equation}
A_{\rm K}\;  =\;  0.114\; A_{\rm V}, 
\end{equation}
where $d$ is the distance in pc, and $A_{\rm V}$ and $A_{\rm K}$ are
the interstellar absorptions in the $V$ and $K$ bands, respectively. 
We have adopted the extinction
model of \citet{Hakkila-1997:a} with the $A_{\rm K}/A_{\rm V}$ ratio
from \citet{Cardelli-1989:a}.  We do not attempt
to include circumstellar extinction, since this is shown by infrared color
to be small or non-existent for the R stars.  The extinction
corrections to the $K$ magnitudes are small, less than 0.2 mag.

\begin{figure}
\resizebox{\hsize}{!}{\includegraphics{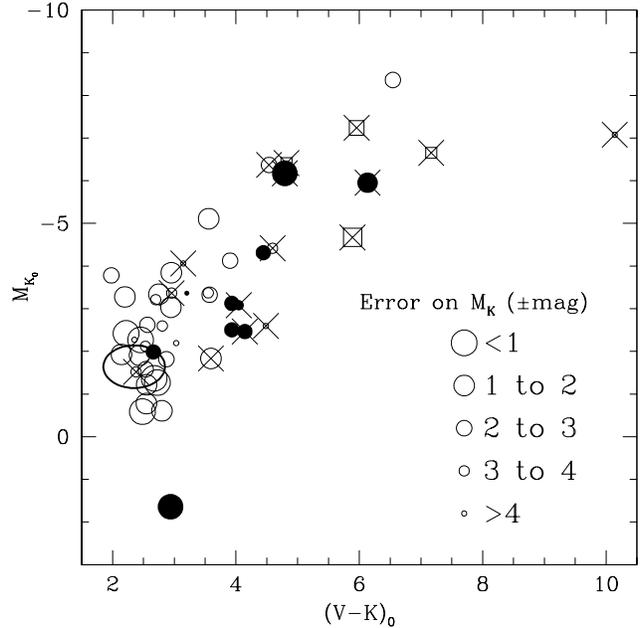}}
\caption{
Dereddened absolute magnitude $M_{\rm K_0}$ versus $(V - K)_0$ 
color for R stars.  
Open symbols: stars of spectral type R0 
-- R3; filled symbols: stars of type 
R4 -- R8; squares: stars of unknown R subtype.
The symbol size is inversely proportional to 
the error on $M_{\rm K}$, as calculated
from the upper and lower bounds on the parallax listed in
Tables~\ref{Tab:dataearly} \& \ref{Tab:datalate}. 
Large crosses denote stars with
$H-K > 0.3$ that are more likely to be N-type rather than R-type
carbon stars (see Fig.~\ref{Fig:JHK}). Only stars with $K \le 8$ 
have been included. The ellipse indicates the
average location of clump giants, according to \protect\citet[][see also
Fig.~\protect\ref{fig2}]{Alves-2000:a}
}
\label{fig1}
\end{figure}

Fig.~\ref{fig1} 
shows the color -- absolute magnitude diagram, $M_{{\rm K}_0}$ versus $(V-K)_0$,
for the stars with available $K$ magnitudes.
The error bars on $M_{\rm K}$ are
calculated from the upper and lower bounds on the parallax 
listed in Tables~\ref{Tab:dataearly} \& \ref{Tab:datalate}.  As noted in Paper~I, the parallax
uncertainties are computed so as to produce symmetric error bars on the 
magnitude.
The resulting uncertainties
in the absolute magnitudes are often larger than 1 mag (see
Fig.~\ref{fig1}), 
and dwarf the 
uncertainties on the measured apparent magnitudes and on the interstellar 
extinction.

Fig.~\ref{fig1} shows that the difference between late- and early-type
R stars, assessed from their photometric, spectral, variability and mass-loss
properties as discussed in Sect.~\ref{Sect:sptype}, is also apparent in their
luminosities:  
late-type R stars with $(V-K)_0 > 4$ tend to be more luminous than 
early-type R stars. Late-type R stars thus clearly mark the bottom of the
asymptotic giant branch.

\subsection{Average absolute magnitude of early-type R stars}
\label{Sect:subaverage}

To derive an average absolute magnitude from a set of measured parallaxes is an
operation plagued by many biases \citep[e.g.,][]{Arenou-1999:a}, 
and several simulations have been performed to identify their importance for the
present sample.

\begin{figure}
\resizebox{\hsize}{!}{\includegraphics{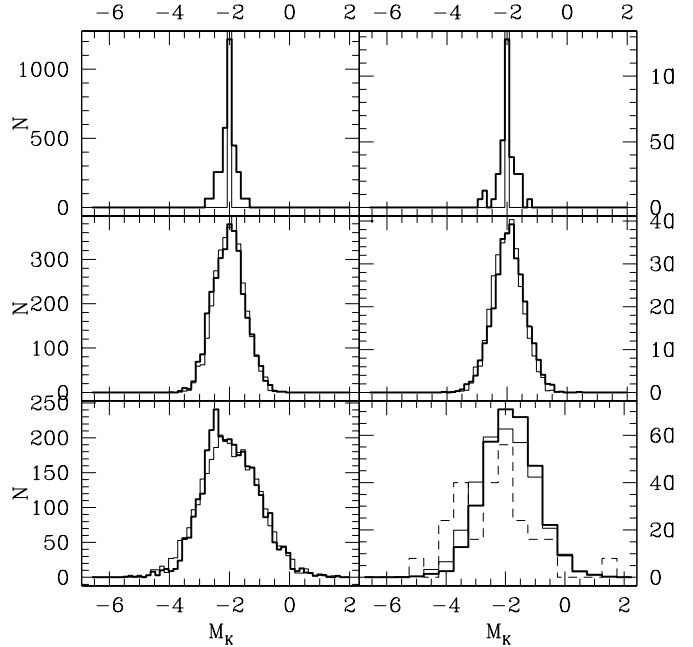}}
\caption[]{\label{Fig:MK_hist}The distribution of absolute magnitudes 
$M_{\rm K}$ for the sample of R stars with $K\le8$ and $(V-K)_0 < 4$ 
obtained by reprocessing the IAD as indicated in
Sect.~\protect\ref{Sect:paral} (thick line),
for different assumptions of the true $M_{\rm K}$ distribution (thin
line): upper panel: $M_K^0
=  -2.0$ for all stars; (ii) middle panel: Gaussian distribution of mean $M_K^0
=  -2.0$ and $\sigma_0 = 0.5$~mag; (iii) bottom panel: Gaussian distribution of
mean $M_K^0 =  -2.0$ and $\sigma_0 = 1.0$~mag. The left panels correspond to the
simulations performed on the IAD files with abscissa residuals drawn from a
normal distribution of mean zero and $\sigma_v = 1$~mas, while the right panels
correspond to the simulations performed on the IAD files with the actual
abscissa residuals. 
In the bottom right panel, the actual {\it observed}
distribution (scaled by a factor of 30) is represented by the
dashed line. The bin size in that panel has been enlarged due to
the smaller size of the observed star sample
}
\end{figure}

More precisely, the simulations described below evaluate  whether the reduction
scheme used in the present
work is able to correctly retrieve the absolute magnitude of the sample stars.
In a first step, it is assumed that all the 59 stars from the present
sample with a $K$ magnitude available have the same
absolute magnitude, namely $M^0_{\rm K} = -2.0$. In that framework, a true
parallax may be 
assigned to each star of the sample, depending on its observed apparent $K$
magnitude according to the relation $\log \varpi = 0.2 (M^0_{\rm K} - K)
- 1.$  
In the first simulation, the original abscissa residuals of the IAD file were
replaced by residuals drawn from a normal distribution of zero mean and
$\sigma_v = 1$~mas (standard deviation of the residuals). To artificially 
increase the sample size, 32 different sets of residuals are drawn for the IAD 
file corresponding to each 
of the 59 R stars, totaling to 1888 simulated stars. In this test, the IAD
file only serves to define the time sampling of the observations and the 
scanning angle on the sky for each observation.  
The  reduction scheme outlined in Sect.~\ref{Sect:paral} is then
applied to the total sets of modified IAD files. Results are shown in
the upper left panel of Fig.~\ref{Fig:MK_hist}. In a second
simulation, the reduction scheme 
is applied to the actual abscissa residuals which  do not necessarily follow a normal
distribution with a standard deviation of $\sigma_v = 1$~mas. Any differences
between the results of the first and second simulation will thus
highlight the impact of the measurement errors. 

\begin{table*}[htb]
\caption[]{\label{Tab:MK_hist}
Weighted average, standard deviation and coefficient of skewness 
of the retrieved $M_K$ distributions 
for different choices of the `true' ($\overline{M_K^0}, \sigma_0)$ 
distributions and input
IAD files (see text)}
\begin{tabular}{rrrrrrrrr}
\noalign{a. All early R stars:}\\
\hline
IAD abscissa residuals: & \multicolumn{3}{c}{Gaussian with
$\sigma_v = 1$~mas}
&&\multicolumn{3}{c}{actual}\\
\cline{2-4}\cline{6-8}\\
           & $\overline{M_K}$  & $\sigma$ & skewness  && $\overline{M_K}$ & $\sigma$ & skewness \\
\hline
\multicolumn{1}{l}{True distribution:}\\
$\overline{M_K^0} = -2.0, \sigma_0 = 0.0$ mag & $-2.01$ & 0.29 & $0.09$
&& $-1.61$ & 1.24 & $3.15$\\  
$\overline{M_K^0} = -2.0, \sigma_0 = 0.5$ mag & $-1.98$ & 0.54 & $0.01$
&& $-1.53$ & 1.29 & $2.59$\\ 
$\overline{M_K^0} = -2.0, \sigma_0 = 1.0$ mag & $-1.89$ & 0.91 & $0.16$
&& $-1.45$ & 1.53 & $1.63$ \\ \hline
\medskip\\
\noalign{b. Only early R stars with $K < 8$:}\\
\hline
IAD abscissa residuals: & \multicolumn{3}{c}{Gaussian with
$\sigma_v = 1$~mas}
&&\multicolumn{3}{c}{actual}\\
\cline{2-4}\cline{6-8}\\
           & $\overline{M_K}$  & $\sigma$ & skewness  && $\overline{M_K}$ & $\sigma$ & skewness \\
\hline
\multicolumn{1}{l}{True distribution:}\\
$\overline{M_K^0} = -2.0, \sigma_0 = 0.0$ mag & $-2.05$ & 0.24 & $-0.24$
&& $-2.02$ & 0.29 & $-0.58$\\  
$\overline{M_K^0} = -2.0, \sigma_0 = 0.5$ mag & $-2.03$ & 0.53 & $0.07$
&& $-1.93$ & 0.54 & $-0.05$\\ 
$\overline{M_K^0} = -2.0, \sigma_0 = 1.0$ mag & $-1.95$ & 0.93 & $0.20$
&& $-1.84$ & 0.90 & $0.03$ \\ \hline
\hline
\end{tabular}
\end{table*}

The results are shown in the upper panel of Fig.~\ref{Fig:MK_hist}, and the
first three moments of the retrieved distributions are listed
in Table~\ref{Tab:MK_hist}, where the coefficient of skewness is
defined as $\mu_3/\sigma^3$ (and $\mu_3$ is the centered moment of
order 3). The method is able to retrieve the true absolute
magnitude when the uncertainties on the residuals have a standard deviation of
1~mas. If the actual residuals are used instead, the computed distribution is
retrieved as well, except for a tail extending towards fainter $M_{\rm
  K}$, as apparent on Fig.~\ref{Fig:MK_bias}. That tail is not visible
in Fig.~\ref{Fig:MK_hist}, though, since stars with $K > 8$ have 
very large and noisy abscissa residuals and  were not
displayed.

Other simulations were carried out as above, except that Gaussian distributions
with $\sigma_0 = 0.5$ and 1~mag were adopted for the true absolute magnitude
distribution.
Basically the same features as for the case $\sigma_0 = 0$ emerge, namely  the
fact that the retrieved distribution using the actual IAD residuals 
is skewed towards fainter
absolute magnitudes, with
a bias of 0.55~mag for the $\sigma_0 = 1$~mag case, which however
reduces to 0.16~mag if only stars with $K \le 8$ are kept (Table~\ref{Tab:MK_hist}).

The distribution derived from the true distribution  
($\overline{M_K^0}, \sigma_0) = (-2.0, 1.0)$ 
provides a fair match to the observed distribution (bottom right panel of
Fig.~\ref{Fig:MK_hist}),
characterized by a weighted-average absolute magnitude 
\begin{equation}
\overline{M_{\rm K}} = \frac{\Sigma_{i = 1}^N M_{{\rm K},i} w_i}{\Sigma_{i =
1}^N w_i} = -1.54,  
\end{equation}
where  $M_{{\rm K},i}$ is the individual absolute magnitude of R stars with
$(V-K)_0 < 4$ and $K \le 8$,
$w_i = 1/\epsilon(M_{{\rm K},i})^2$ and $\epsilon(M_{{\rm K},i}) = 5 \;
(\log
\varpi_U - \log \varpi_0) =  5 \; (\log \varpi_0 - \log \varpi_L)$, and   
$\varpi_0, \varpi_L, \varpi_U$ as derived by our reduction scheme are given in
Tables~\ref{Tab:dataearly} \& \ref{Tab:datalate}. 

Thus it may be concluded that the early-type R stars have a true 
absolute-magnitude distribution close to   
($\overline{M_K^0}, \sigma_0) = (-2.0, 1.0)$.
Since the present sample of R stars is basically magnitude-limited, it 
is subject to the Malmquist bias favoring high-luminosity stars. The
\citet{Malmquist-1936:a} bias is probably responsible for the high-luminosity
tail and the deficit of low-luminosity stars
in the observed sample (bottom-right panel of
Fig.~\ref{Fig:MK_hist}), as compared to the simulated Gaussian population.

\subsection{Early R stars as members of the red clump}

\begin{figure}
\resizebox{\hsize}{!}{\includegraphics{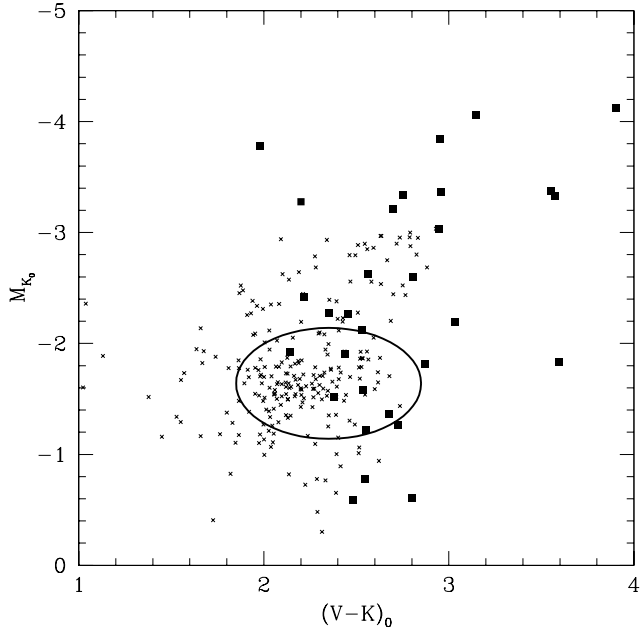}}
\caption{$M_{\rm K}$ vs. $V-K$ for early R stars with $K\le8$ (filled symbols), and red clump giants
(squares) from \cite{Alves-2000:a}. The ellipse is the same as in
Fig.~\protect\ref{fig1}
}
\label{fig2}
\end{figure}

Fig.~\ref{fig2} reproduces the color-magnitude diagram from Fig.~\ref{fig1} for
R stars with $(V - K)_0 < 4$ and $K \le 8$, 
and compares them to 238 nearby red clump stars as compiled by
\cite{Alves-2000:a}. The error bars have been excluded
for clarity.  Red clump stars have become of renewed interest recently 
because Hipparcos has made possible the calibration of their absolute
magnitudes via observations of nearby stars \citep{Paczynski-1998:a,Girardi-1998:a},
making them a powerful tool for distance measurements within the Galaxy and 
the Local Group.  Physically, these objects are the analogues of the red
horizontal branch stars seen in metal-rich globular clusters: helium core burning post-red-giant branch stars.
Fig.~\ref{fig2} very strongly suggests that the R stars 
are members of the red clump, since the inferred true absolute magnitude $M_K^0
= -2.0$ (with an estimated intrinsic dispersion of 1.0 mag) of R stars (Sect.~\ref{Sect:subaverage}) is similar to the average
$M_K = -1.61\pm 0.03$ derived from the sample of \citet{Alves-2000:a}. 
The $V-K$ colors of R stars show a somewhat broader spread (but this may be due
to
contamination by a few N stars), and they are on average
about 0.5~mag redder than the local clump stars, which have $< V -
K > = 2.35$; this is likely
to be due to the presence of weak to moderate $\rm C_2$ and CN band absorption
in the blue region of the spectrum.

\section{The space density of early R stars}

\begin{figure}
\resizebox{\hsize}{!}{\includegraphics{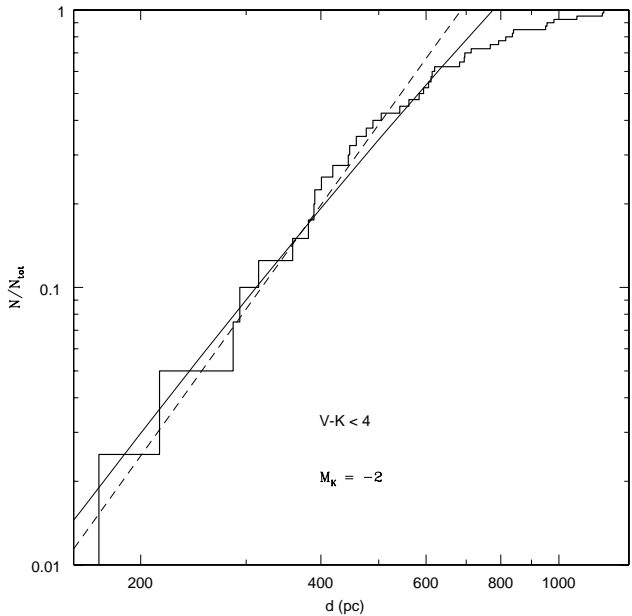}}
\caption{Cumulative frequency distribution of the distances of early R stars
(thick line), as compared to predictions for a stellar sample uniformly
distributed in the plane and with an exponential distribution perpendicular to
the plane (thin line). A scale height of 300~pc and a true  absolute
magnitude $M_K^0 = -2.0$ have been adopted. The thin dashed line corresponds to
the predictions for a spherically-symmetric homogeneous sample
}
\label{Fig:space}
\end{figure}

In this section, the space density of R stars will be estimated and compared to
that of red clump stars. To evaluate the completeness of the sample of early R
stars, Fig.~\ref{Fig:space} displays the cumulative frequency distribution of
early R stars as a function of distance. To circumvent the bias on the derived
absolute magnitude distribution (see Fig.~\ref{Fig:MK_hist}), the distances
have been estimated by assigning the absolute magnitude $M_K^0 = -2.0$ to all
early R stars. Under the assumption of a uniform density $n_0$ in the plane
and an
exponential distribution perpendicular to the plane with scale height $z_0$, the
cumulative frequency distribution varies with distance $d$ as
\begin{displaymath}
\log N(d) = 4\; \pi n_0\; d^3 \times 
\end{displaymath}
\begin{equation}
\left[ \left(\frac{1}{2} \frac{z_0}{d} -
\left(\frac{z_0}{d}\right)^3\right) + \left( \left(\frac{z_0}{d}\right)^2 
+\left(\frac{z_0}{d}\right)^3\right) \exp\left(-\frac{d}{z_0}\right)\right]   
\end{equation}

The scale height and 
the space density of R stars can then be estimated 
from the values that must be assigned to the parameters $z_0$ and 
$n_0$ in order to fit the observed
distribution. It is found to be about $n_0 \sim 4.5\; 10^{-8}
\; {\rm pc}^{-3}$ for $z_0 = 300$~pc. Fig.~\ref{Fig:space} reveals that, for 
this parameter set, early R stars follow the predicted
uniform/exponential distribution up to about 600~pc (involving about 60\% 
of the total sample). It needs
not be emphasized that these values are very uncertain though, as $z_0$ is
only poorly constrained by the fit and both $z_0$ and $n_0$ depend on the
quite uncertain $M_K^0$ assignment.  
Adopting $M_K^0 = -1$ instead yields $n_0 \sim
1.5\;10^{-7}$~pc$^{-3}$ and $z_0 \sim 180$~pc.

The sample of red clump stars by \citet{Alves-2000:a} contains 284 stars
with 5\% parallaxes from the Hipparcos catalogue and is roughly complete
to 12 mas or about 83 pc.  The volume density of the red clump stars
is thus about $\rm 1.14 \; 10^{-4} ~ pc^{-3}$ and the R stars form 
about 0.04\% (or 0.13\% if $M_K^0 = -1$) of the total.  
There are additional uncertainties because,
for a given star, there is no way to tell whether it is a clump star or is
on the red giant branch.  Nevertheless, the qualitative conclusion
is clear: a very small fraction of clump stars in the solar neighborhood
is carbon-rich.

\section{Conclusions}
\label{Sect:conclusions}

In this paper, we have re-analyzed the Hipparcos {\it Inter\-mediate Astrometric
Data} for a sample of 83 carbon stars of class R, the class of warm
evolved carbon red giants. A special reprocessing scheme has been used, fitting
the distance modulus instead of the parallax (thus ensuring its positiveness)
and rejecting outliers within a bootstrap scheme ensuring consistency between
the solutions obtained from the NDAC and FAST data. 
This reduction scheme is not free from biases: Monte-Carlo simulations have
shown that it tends to produce parallaxes that are too large when
$\epsilon(\varpi)/\varpi \ga 1$. Despite these biases, the simulations allow
the identification of the true properties of the population. 

We have shown that the early- and late-type R stars have different properties.
Late-type R stars generally have subtypes later than R4, although a better
definition uses their color indices $(V-K)_0 > 4$ or $(H-K) > 0.3$.  Most of
these stars are long-period variables of the SR or Mira types, and they often
exhibit excess emission at 12~$\mu$m due to dust, indicative of mass loss.  In
the HR diagram, these stars mark the bottom of the AGB, and are thus
related, or even identical, to the N stars.

The early-type R stars form a rather homogeneous group, with most of its members
populating the red clump. They are well represented by a population having a
Gaussian distribution of absolute magnitudes such that $(M_K^0,\sigma_0) = (-2., 1.)$. They are not variable and do not appear to suffer from substantial
mass loss, as judged from the absence of 12~$\mu$m excess. They represent fewer
than 0.1\% of the clump stars. Their nature as carbon stars is not due to low
metallicity: R stars have metal abundances which are close to solar 
\citep{Dominy-1984:a}, except for the carbon excess.

The location of the R 
stars in the red clump provides strong support for the
formation scenario suggested by \citet{Dominy-1984:a}: the helium core 
flash mixes carbon to the surface of the star. 
This mixing episode may perhaps be rotationally-induced
  as the star is spun up by the accretion
  of a former companion. This scenario has the
  advantage of accounting as well for the lack of binary systems among 
  R stars \citep{Mcclure-1997:b}, but it requires a braking mechanism
  (e.g. magnetic braking as advocated by \citet{Habets-1989}?)
  since R stars do not currently appear to be rapid rotators. 
 The fraction of stars which become R stars
is much smaller than the fraction \citep[10\% -- 50\%, see][]{Wallerstein-1998:a} of stars which become carbon stars
on the AGB, and it is unlikely to have any significant influence on the 
number of carbon AGB stars.  It suggests that a small fraction of the 
RR~Lyrae stars and blue horizontal branch stars in the Galactic halo are 
also carbon rich -- these stars however are too hot to have $\rm C_2$
in the atmosphere.  

\begin{acknowledgements}
We thank Bohdan Paczy\'nski and Jim Gunn for useful discussions and advice,
David Alves for a machine-readable copy of his red clump star data,
and the National Aeronautics and Space Administration NASA
who supported this work via grants NAG5-6734 and NAG5-8083. We also thank
Fr\'ed\'eric Arenou, the referee, for his useful comments and suggestions. 
This publication makes use of data products from the {\it Two Micron All Sky
Survey}, which is a joint project of the University of Massachusetts and the
Infrared Processing and Analysis Center/California Institute of
Technology, funded by the National Aeronautics and Space Administration
and the National Science Foundation.  This research has made use of the 
Simbad data base, operating at CDS, Strasbourg, France.

\end{acknowledgements}

\bibliographystyle{apj}
\bibliography{articles,books}

\begin{thebibliography}{63}
\expandafter\ifx\csname natexlab\endcsname\relax\def\natexlab#1{#1}\fi

\bibitem[{{Alves}(2000)}]{Alves-2000:a}
{Alves}, D.~R. 2000, ApJ, 539, 732

\bibitem[{{Arenou} \& {Luri}(1999)}]{Arenou-1999:a}
{Arenou}, F. \& {Luri}, X. 1999, in Harmonizing Cosmic Distance Scales in a
  Post-Hipparcos Era ASP Conference Series \#167, ed. D.~{Egret} \& A.~{Heck}

\bibitem[{{Barnbaum} {et~al.}(1996){Barnbaum}, {Stone}, \&
  {Keenan}}]{Barnbaum-1996:a}
{Barnbaum}, C., {Stone}, R. P.~S., \& {Keenan}, P.~C. 1996, ApJS, 105, 419

\bibitem[{{Bartkevi\v{c}ius}(1996)}]{Bartkevicius-1996:a}
{Bartkevi\v{c}ius}, A. 1996, Baltic Astronomy, 5, 217

\bibitem[{{Boffin} {et~al.}(1993){Boffin}, {Abia}, {Isern}, \&
  {Rebolo}}]{Boffin-1993:b}
{Boffin}, H. M.~J., {Abia}, C., {Isern}, J., \& {Rebolo}, R. 1993, A\&AS, 102,
  361

\bibitem[{{Bond}(1975)}]{Bond-1975:a}
{Bond}, H.~E. 1975, ApJ, 202, L47

\bibitem[{{Cannon}(1921)}]{Cannon-1921:a}
{Cannon}, A.~J. 1921, Circ. Harvard Obs., 224

\bibitem[{{Cardelli} {et~al.}(1989){Cardelli}, {Clayton}, \&
  {Mathis}}]{Cardelli-1989:a}
{Cardelli}, J.~A., {Clayton}, G.~C., \& {Mathis}, J.~S. 1989, ApJ, 345, 245

\bibitem[{{Catchpole} {et~al.}(1979){Catchpole}, {Robertson}, {Lloyd Evans},
  {Feast}, {Glass}, \& {Carter}}]{Catchpole-1979:a}
{Catchpole}, R.~W., {Robertson}, B. S.~C., {Lloyd Evans}, T. H.~H., {Feast},
  M.~W., {Glass}, I.~S., \& {Carter}, B.~S. 1979, S.~Afr.~Obs.~Circ., 1, 61

\bibitem[{{Cohen} {et~al.}(1981){Cohen}, {Frogel}, {Persson}, \&
  {Elias}}]{Cohen-1981:a}
{Cohen}, J.~G., {Frogel}, J.~A., {Persson}, S.~E., \& {Elias}, J.~H. 1981, ApJ,
  249, 481

\bibitem[{{Cottrell} \& {Lawson}(1998)}]{Cottrell-1998:a}
{Cottrell}, P.~L. \& {Lawson}, W.~A. 1998, Publ. Astron. Soc. Aust., 15, 179

\bibitem[{{Dahn} {et~al.}(1977){Dahn}, {Liebert}, {Kron}, {Spinrad}, \&
  {Hintzen}}]{Dahn-1977:a}
{Dahn}, C.~C., {Liebert}, J., {Kron}, R.~G., {Spinrad}, H., \& {Hintzen}, P.~M.
  1977, ApJ, 216, 757

\bibitem[{{Dean}(1976)}]{Dean-1976:a}
{Dean}, C.~A. 1976, AJ, 81, 364

\bibitem[{{Dominy}(1984)}]{Dominy-1984:a}
{Dominy}, J.~F. 1984, ApJS, 55, 27

\bibitem[{{Dominy} {et~al.}(1986){Dominy}, {Lambert}, {Gehrz}, \&
  {Mozurkewich}}]{Dominy-1986:a}
{Dominy}, J.~F., {Lambert}, D.~L., {Gehrz}, R.~D., \& {Mozurkewich}, D. 1986,
  AJ, 91, 951

\bibitem[{{Elias}(1978)}]{Elias-1978:a}
{Elias}, J.~H. 1978, AJ, 83, 791

\bibitem[{{ESA}(1997)}]{Hipparcos}
{ESA}. 1997, The Hipparcos and Tycho Catalogues (ESA SP-1200)

\bibitem[{{Feast} {et~al.}(1997){Feast}, {Carter}, {Roberts}, {Marang}, \&
  {Catchpole}}]{Feast-1997:a}
{Feast}, M.~W., {Carter}, B.~S., {Roberts}, G., {Marang}, F., \& {Catchpole},
  R.~M. 1997, MNRAS, 285, 317

\bibitem[{{Feast} \& {Glass}(1973)}]{Feast-1973:a}
{Feast}, M.~W. \& {Glass}, I.~S. 1973, MNRAS, 161, 293

\bibitem[{{Feast} \& {Whitelock}(1992)}]{Feast-1992:a}
{Feast}, M.~W. \& {Whitelock}, P.~A. 1992, MNRAS, 259, 6

\bibitem[{{Gezari} {et~al.}(1999){Gezari}, {Pitts}, \& {Schmitz}}]{CaInOb}
{Gezari}, D.~Y., {Pitts}, P.~S., \& {Schmitz}, M. 1999, Catalog of Infrared
  Observations, 5th edn.

\bibitem[{{Girardi} {et~al.}(1998){Girardi}, {Giuricin}, {Mardirossian},
  {Mezzetti}, \& {Boschin}}]{Girardi-1998:a}
{Girardi}, M., {Giuricin}, G., {Mardirossian}, F., {Mezzetti}, M., \&
  {Boschin}, W. 1998, ApJ, 505, 74

\bibitem[{{Green} \& {Margon}(1994)}]{Green-1994:a}
{Green}, P.~J. \& {Margon}, B. 1994, ApJ, 423, 723

\bibitem[{{Habets} \& {Zwaan}(1989)}]{Habets-1989}
{Habets}, G. M. H.~J. \& {Zwaan}, C. 1989, A\&A, 211, 56

\bibitem[{{Hakkila} {et~al.}(1997){Hakkila}, {Myers}, {Stidham}, \&
  {Hartmann}}]{Hakkila-1997:a}
{Hakkila}, J., {Myers}, J.~M., {Stidham}, B.~J., \& {Hartmann}, D.~H. 1997, AJ,
  114, 2042

\bibitem[{{Iben} \& {Renzini}(1983)}]{Iben-1983:a}
{Iben}, Jr., I. \& {Renzini}, A. 1983, ARA\&A, 21, 271

\bibitem[{{Jorissen} \& {Knapp}(1998)}]{Jorissen-1998:b}
{Jorissen}, A. \& {Knapp}, G.~R. 1998, A\&AS, 129, 363

\bibitem[{{Jorissen} {et~al.}(1998){Jorissen}, {{V}an {E}ck}, {Mayor}, \&
  {Udry}}]{Jorissen-1998:a}
{Jorissen}, A., {{V}an {E}ck}, S., {Mayor}, M., \& {Udry}, S. 1998, A\&A, 332,
  877

\bibitem[{{Keenan}(1942)}]{Keenan-1942:a}
{Keenan}, P.~C. 1942, ApJ, 96, 101

\bibitem[{{Keenan}(1993)}]{Keenan-1993:a}
---. 1993, PASP, 105, 905

\bibitem[{{Kerschbaum} \& {Hron}(1994)}]{Kerschbaum-1994:a}
{Kerschbaum}, F. \& {Hron}, J. 1994, A\&AS, 106, 397

\bibitem[{{Kholopov} {et~al.}(1998){Kholopov}, {Samus}, {Frolov}, {Goranskij},
  {Gorynya}, {Karitskaya}, {Kazarovets}, {Kireeva}, {Kukarkina}, {Kurochkin},
  {Medvedeva}, {Pastukhova}, {Perova}, {Rastorguev}, \&
  {Shugarov}}]{Kholopov-1998:a}
{Kholopov}, P.~N., {Samus}, N.~N., {Frolov}, M.~S., {Goranskij}, V.~P.,
  {Gorynya}, N.~A., {Karitskaya}, E.~A., {Kazarovets}, E.~V., {Kireeva}, N.~N.,
  {Kukarkina}, N.~P., {Kurochkin}, N.~E., {Medvedeva}, G.~I., {Pastukhova},
  E.~N., {Perova}, N.~B., {Rastorguev}, A.~S., \& {Shugarov}, S.~Y. 1998,
  Combined General Catalogue of Variable Stars, 4th edn.

\bibitem[{{Knapp}(2001)}]{Knapp-2001:a}
{Knapp}, G.~R. 2001, ApJ, (submitted)

\bibitem[{{Lee} {et~al.}(1944){Lee}, {Gore}, \& {Bartlett}}]{Lee-1944:a}
{Lee}, O.~J., {Gore}, G.~D., \& {Bartlett}, T.~J. 1944, Annals Dearborn Obs.,
  5, 1

\bibitem[{{Malmquist}(1936)}]{Malmquist-1936:a}
{Malmquist}, K.~G. 1936, Meddel. Stockholm Obs., 26

\bibitem[{{Mayall}(1951)}]{Mayall-1951:a}
{Mayall}, M.~M. 1951, Bull. Harvard Obs., 920, 32

\bibitem[{{Mc{C}lure}(1984{\natexlab{a}})}]{Mcclure-1984:a}
{Mc{C}lure}, R.~D. 1984{\natexlab{a}}, PASP, 96, 117

\bibitem[{{Mc{C}lure}(1984{\natexlab{b}})}]{Mcclure-1984:b}
---. 1984{\natexlab{b}}, ApJ, 280, L31

\bibitem[{{Mc{C}lure}(1997{\natexlab{a}})}]{Mcclure-1997:a}
---. 1997{\natexlab{a}}, PASP, 109, 536

\bibitem[{{Mc{C}lure}(1997{\natexlab{b}})}]{Mcclure-1997:b}
---. 1997{\natexlab{b}}, PASP, 109, 256

\bibitem[{{Mendoza V} \& {Johnson}(1965)}]{Mendoza-1965:a}
{Mendoza V}, E.~E. \& {Johnson}, H.~L. 1965, ApJ, 141, 161

\bibitem[{{Neugebauer} \& {Leighton}(1969)}]{IRC}
{Neugebauer}, G. \& {Leighton}, R.~B. 1969, Two-Micro Sky Survey Catalogue
  (NASA SP-3047)

\bibitem[{{Noguchi} {et~al.}(1981){Noguchi}, {Kawara}, {Kobayashi}, {Okuda},
  {Sato}, \& {Oishi}}]{Noguchi-1981:a}
{Noguchi}, K., {Kawara}, K., {Kobayashi}, Y., {Okuda}, H., {Sato}, S., \&
  {Oishi}, M. 1981, PASJ, 33, 373

\bibitem[{{Ohnaka} \& {Tsuji}(1999)}]{Ohnaka-1999:a}
{Ohnaka}, K. \& {Tsuji}, T. 1999, A\&A, 345, 233

\bibitem[{{Paczy\'nski} \& {Stanek}(1998)}]{Paczynski-1998:a}
{Paczy\'nski}, B. \& {Stanek}, K.~Z. 1998, ApJ, 494, L219

\bibitem[{{Pickering}(1896)}]{pickering-1896:a}
{Pickering}, E.~C. 1896, Harvard Circ

\bibitem[{{Pickering}(1908)}]{Pickering-1908:a}
---. 1908, Harvard Circ

\bibitem[{{Pourbaix} \& {Jorissen}(2000)}]{Pourbaix-2000:b}
{Pourbaix}, D. \& {Jorissen}, A. 2000, A\&AS, 145, 161

\bibitem[{{Pourbaix} {et~al.}(2001{\natexlab{a}}){Pourbaix}, {Knapp}, \&
  {Jorissen}}]{Pourbaix-2001:c}
{Pourbaix}, D., {Knapp}, G.~R., \& {Jorissen}, A. 2001{\natexlab{a}}, A\&A, (in
  preparation)

\bibitem[{{Pourbaix} {et~al.}(2001{\natexlab{b}}){Pourbaix}, {Knapp}, \&
  {Jorissen}}]{Pourbaix-2001:d}
---. 2001{\natexlab{b}}, A\&A, (in preparation)

\bibitem[{{Sanford}(1944)}]{Sanford-1944:a}
{Sanford}, R.~F. 1944, ApJ, 99, 145

\bibitem[{{Scalo}(1976)}]{Scalo-1976:a}
{Scalo}, J.~M. 1976, ApJ, 206, 474

\bibitem[{{Shane}(1928)}]{Shane-1928:a}
{Shane}, C.~D. 1928, Lick Obs. Bull., 13, 123

\bibitem[{{Skrutskie}(1997)}]{Skrutskie-1997:a}
{Skrutskie}, M. 1997, S\&T, 94, 46

\bibitem[{{Slettebak} {et~al.}(1969){Slettebak}, {Keenan}, \&
  {Brundage}}]{Slettebak-1969:a}
{Slettebak}, A., {Keenan}, P.~C., \& {Brundage}, R.~K. 1969, AJ, 74, 373

\bibitem[{{Stephenson}(1989)}]{Stephenson-1989:a}
{Stephenson}, C.~B. 1989, Publ. Warner \& Swasey Obs., 3, 1

\bibitem[{{Stock} \& {Wroblewski}(1972)}]{Stock-1972:a}
{Stock}, J. \& {Wroblewski}, H. 1972, Publ. Dept. Astron. Univ. Chile, 2, 1

\bibitem[{{Ulla} {et~al.}(1997){Ulla}, {Thejll}, {Kipper}, \&
  {Jorgensen}}]{Ulla-1997:a}
{Ulla}, A., {Thejll}, P., {Kipper}, T., \& {Jorgensen}, U.~G. 1997, A\&A, 319,
  244

\bibitem[{{van {L}eeuwen} \& {Evans}(1998)}]{vanLeeuwen-1998:a}
{van {L}eeuwen}, F. \& {Evans}, D.~W. 1998, A\&AS, 130, 157

\bibitem[{{Vandervoort}(1958)}]{Vandervort-1958:a}
{Vandervoort}, G.~L. 1958, AJ, 63, 477

\bibitem[{{Wallerstein} \& {Knapp}(1998)}]{Wallerstein-1998:a}
{Wallerstein}, G. \& {Knapp}, G.~K. 1998, ARA\&A, 36, 369

\bibitem[{{Whitelock} {et~al.}(2000){Whitelock}, {Marang}, \&
  {Feast}}]{Whitelock-2000:a}
{Whitelock}, P., {Marang}, F., \& {Feast}, M. 2000, MNRAS, (submitted)

\bibitem[{{Yamashita}(1975)}]{Yamashita-1975:a}
{Yamashita}, Y. 1975, PASJ, 27, 325

\end{thebibliography}
\end{document}